\documentstyle[12pt]{article}
\newcommand{\la}{\lambda}
\newcommand{\vi}{\varphi}
\newcommand{\mg}{m_\pi}
\newcommand{\ms}{m_\sigma}
\newcommand{\tr}{\mbox{Tr}}
\newcommand{\Tr}{\mbox{\bf Tr}}

       \def\de{depth}

\let\a=\alpha \let\be=\beta  \let\de=\delta
   
  \let\la=\lambda 
    \let\s=\sigma
 
\let\ph=\varphi  \let\PH=\Phi 
\let\Om=\Omega  
  \let\D=\Delta

\def\0{\over } \def\1{\vec }     \def\2{{1\over2}} \def\4{{1\over4}}
\def\5{\bar }  \def\6{\partial } \def\7#1{{#1}\llap{/}}
\def\8#1{{\textstyle{#1}}}       \def\9#1{{\bf {#1}}}
 \def\llp{\hbox to 0pt{\hss /\hskip1.5pt}}
\def\llo{\hbox to 0.2pt{\hss /}} \def\llq{\hbox to 0pt{\hss /\hskip0.5pt}}
\def\so{\supset\hbox to 0pt{\hss $\displaystyle -$\hskip1pt}}

\def\<{\langle } \def\>{\rangle }

   \let\hc=\dagger

\let\nn=\nonumber
\def\bea{\begin{eqnarray}} \def\eea{\end{eqnarray}}
\def\beann{\begin{eqnarray*}} \def\eeann{\end{eqnarray*}}
\def\beq{\begin{equation}} \def\eeq{\end{equation}}

\date{}
\title{
{\large\rm DESY 95-028}\hfill{\large\tt ISSN 0418-9833}\\
{\large\rm February 1995}\hfill\vspace*{3.0cm}\\
Thermodynamics of the Electroweak Phase Transition}
\author{W. Buchm\"{u}ller, Z. Fodor\thanks{Address after 1
January 1995: Theory Division, CERN, CH-Geneva 23,\hspace*{4cm}\mbox{}
\hspace*{.63cm}on leave from Institute for Theoretical Physics,
E\"otv\"os University, Budapest, Hungary}
\ and A. Hebecker\\
{\normalsize\it Deutsches Elektronen-Synchrotron DESY, 22603 Hamburg, Germany}
\vspace*{3.0cm}\\
}
\begin{document}

\setlength{\baselineskip}{18pt}
\maketitle
\begin{abstract}
\thispagestyle{empty}
\noindent
We discuss several general aspects of the free energy of the standard model at
high temperatures. In particular the Clausius-Clapeyron equation is shown to
yield a relation between the latent heat and the jump in the order parameter.
The free energy is calculated as function of temperature in resummed
perturbation theory to two-loop order. A new resummation procedure is proposed
in which the symmetric phase and the Higgs phase are treated differently. A
quantitative description of the phase transition is achieved for Higgs masses
below $\sim 70$ GeV. The results are found to be in agreement with recent
numerical simulations on large lattices. The phase transition provides no
evidence for strong non-perturbative effects in the symmetric phase.
\end{abstract}
\setcounter{page}{0}
\newpage
\section{Introduction}
In the standard model of electroweak interactions all masses are generated by
the Higgs mechanism. At high temperatures this implies, in analogy to
superconductivity, a phase transition from a massive low-temperature phase to a
massless high-temperature phase, where the electroweak symmetry is ``restored''
\cite{kirzh}. This transition is of great cosmological importance because
baryon-number violating processes fall out of thermal equilibrium below the
critical temperature of the phase transition \cite{kuzmin}. As a consequence,
the present value of the baryon asymmetry of the universe has finally been
determined at the electroweak transition.

In recent years quantitative studies of the electroweak phase transition have
been carried out by means of resummed perturbation theory
\cite{dine}-\cite{hebecker} and lattice Monte Carlo simulations
\cite{bunk}-\cite{karsch}. As a first step towards a treatment of the full
standard model, the pure SU(2) Higgs model is usually investigated, neglecting
fermion effects and the mixing between photon and neutral vector boson. These
can be treated in perturbation theory and do not affect the essential infrared
problem. There
is general agreement that the phase transition is first-order for Higgs masses
$m_H$ below the vector boson mass $m_W$. At larger Higgs masses lattice results
\cite{evertz}, general arguments \cite{reuter} and non-perturbative solutions
of gap equations \cite{owe} suggest that the first-order transition changes to
a smooth crossover. However, this conjecture still remains to be firmly
established.

The goal of the present paper is a quantitative study of the thermodynamics
of the transition for Higgs masses below 80 GeV. Our analysis will be based
on the gauge invariant ``order parameter'' $\PH^{\hc}\PH$ \cite{jansen}
and the corresponding free energy \cite{luscher}. We shall extend our
previous one-loop analysis \cite{bfh} to two-loop order and discuss the
connection with the conventional Landau gauge approach. The comparison
between one-loop and two-loop results, and also between different resummation
procedures, will allow us to estimate the uncertainty of perturbative
predictions for thermodynamic observables.

It is well known that the electroweak phase transition is influenced by
non-perturbative effects whose size is governed by the confinement scale of the
effective three-dimensional theory which describes the high-temperature limit
of the SU(2) Higgs model. These effects are particularly relevant in the
symmetric phase, and one may worry to what extent a purely perturbative
analysis of the phase transition can yield sensible results at all. However, it
is not yet known how
important these non-perturbative effects are quantitatively for
different Higgs masses. As we shall see, at two-loop order our results
depend only logarithmically on the infrared cutoff needed in the
symmetric phase. We shall then compare the perturbative results
with recent non-perturbative results obtained by numerical
Monte Carlo simulations on large lattices \cite{montvay}.
This will enable us to estimate the size of possible non-perturbative
corrections.

The paper is organized as follows. In sect. 2 we shall discuss
several general aspects of the free energy and the various effective
potentials used in connection with the electroweak phase transition.
The description of the transition is based on a gauge invariantly coupled
source term. The corresponding gauge invariant free energy can be obtained from
the minima of the usual Landau gauge potential.

In sect. 3 a useful relation is derived between the latent heat
and the gauge invariant order parameter, which follows from the
Clausius-Clapeyron equation by dimensional arguments. This relation is
exactly satisfied in perturbation theory, and it holds for the lattice results
within the estimated errors.

Sect. 4 deals with the free energy at one-loop. A new resummation procedure is
introduced which treats the Higgs phase and the symmetric phase differently,
reproducing the numerical results of \cite{bfh} in a different manner.
This also clarifies the connection between the gauge invariant and the usual
Landau gauge approach. It is shown that the barrier of the gauge invariant
effective potential is given by analytic continuations of the convex
potential defined by a Legendre transformation.

In Sect. 5 higher order corrections are obtained using the new resummation
method of the previous section. This calculation is an
extension of the gauge invariant approach of \cite{bfh} to the two-loop level.
While previous numerical results for phase transition parameters are
qualitatively reproduced, the convergence of perturbation theory is
improved in the new approach.

Sect. 6 contains a comparison with lattice results, showing quantitative
agreement within the statistical and systematic errors. For this the most
important zero-temperature renormalization effects have to be included at small
Higgs masses.

Conclusions are summarized in sect. 7, and the appendix contains the explicit
formulae used in sect. 5.

\section{Free energy of the SU(2) Higgs model}

The action of the SU(2) Higgs model at finite temperature $T$ reads
\beq \label{ST}
S_{\beta}[\PH,W] = \int_{\beta} dx \; \tr \left[
{1\over 2}W_{\mu\nu}W_{\mu\nu} +
(D_{\mu}\PH)^\hc D_{\mu}\PH + \mu \PH^\hc \PH
+ 2 \lambda (\PH^\hc \PH)^2 \right] \, ,
\eeq
with
\bea
\PH &=& \2 (\s + i \vec{\pi}\cdot \vec{\tau}) \, ,\quad
D_{\mu}\PH = (\6_{\mu} - i g W_{\mu})\PH\, ,\quad
W_{\mu} = \2\vec{\tau}\cdot \vec{W_{\mu}}\ ,\\
&&\quad \int_{\beta}dx = \int_0^{\beta}d\tau\int_{\Om}d^3x\ ,\quad
\beta = {1\over T}\ .
\eea
Here $\vec{W_{\mu}}$ is the vector field, $\s$ is the Higgs field, $\vec{\pi}$
is the Goldstone field and $\vec{\tau}$ is the triplet of Pauli matrices.
In general, we shall consider the limit of infinite
spatial volume $\Om$. For perturbative
calculations gauge fixing and ghost terms have to be added to the action
(\ref{ST}).

The free energy density of the system, $W(T,J)$, is given by the partition
function, i.e., the trace of the density matrix,
\beq
\exp(-\be\Om W(T,J)) = \Tr \exp\left[-\be\left(\hat{H}+
J \int_{\Om}d^3x \hat{\PH}^{\hc}\hat{\PH}\right)\right]\ ,
\eeq
where $\hat{H}$ is the Hamilton operator of the theory, and $\hat{\PH}$ is the
operator describing the Higgs field. We have added a source $J$, with
$\6_{\mu}J = 0$, coupled to the spatial average of the gauge invariant
composite operator $\hat{\PH}^{\hc}\hat{\PH}$ (here and below the trace
operator acting on $\hat{\PH}^{\hc}\hat{\PH}$ is omitted for brevity).
Similarly, one may define a free energy for spatially varying sources $J(x)$.
The partition function can be expressed as a euclidian functional integral
\cite{kapusta},
\beq \label{FI}
\exp\left(-\be\Om W(T,J)\right) = \int_{\be}D\PH D\PH^{\hc} DW_{\mu}
\exp\left(-\int_{\be}dx\left(L + J \PH^{\hc}\PH\right)\right)\ ,
\eeq
where {\it L} is the euclidean lagrangian density, and
the bosonic fields $\PH$ and $W_{\mu}$ satisfy periodic boundary
conditions at $\tau =0$ and $\tau = \be$. Eq. (\ref{FI}) is the
starting point of perturbative as well as numerical evaluations of the free
energy.

Note, that the source $J$ in eq. (\ref{FI}) couples to a gauge invariant
composite field.  Hence, the free energy $W(T,J)$ is gauge independent.
The spatially constant source $J$ simply redefines the mass term in the
action (\ref{ST}).
This is in contrast to the usually considered generating function of
connected Green functions at zero momentum,
\beq
\exp\left(-\be\Om \tilde{W}(T,j;J)\right) =
\int_{\be}D\PH D\PH^{\hc} DW_{\mu}\exp\left(-\int_{\be}dx\left(L +
J \PH^{\hc}\PH + j\s \right)\right)\ .
\eeq
Here the source $j$ couples to a gauge dependent quantity, the field $\s$.
Consequently, $\tilde{W}(T,j;J)$ is gauge dependent and not a
physical observable. For later use we have also kept the dependence on the
source $J$.

The generating function $\tilde{W}(T,j;0)$ can be made finite in the usual way
by a multiplicative renormalization of couplings and fields. This is not
the case for the free energy $W(T,J)$, since $J$ couples to a composite
field. It is known \cite{zinn1} that two more counter terms,
linear and quadratic in $J$, are necessary
in order to subtract the additional
divergencies. Hence, the renormalized free energy $W(T,J)$ contains two
arbitrary constants in addition to the usual renormalized
parameters at zero temperature.

{}From the free energy $W(T,J)$ a gauge invariant effective potential
$V(T,\rho)$ can be obtained as usual by means of a Legendre transformation,
\beq
V(T,\rho) = W(T,J) - \2 \rho J \ ,
\eeq
where
\beq
\2 \rho \equiv {1\over \Om}\int_{\Om}d^3x <\hat{\PH}^{\hc}(x)\hat{\PH}(x)>
= {\6\over \6 J}W(T,J)
\eeq
is the spatial average of the thermal expectation value of
$\hat{\PH}^{\hc}\hat{\PH}$, which plays the role of an ``order parameter'' in
the SU(2) Higgs model. By definition, the effective potential $V(T,\rho)$ is
convex. The ground state of the theory corresponds to a stationary point of
the effective potential, where $\6 V(T,\rho)/\6\rho$ vanishes. In the case of
a first-order phase transition two stationary points connected by a straight
line (see fig. 3 in sect. 4) represent two coexisting phases.

What is the effect of the ambiguity of the renormalized free energy $W(T,J)$ on
the effective potential? Consider two definitions of the free energy, related
by
\beq
\bar{W}(T,J) = W(T,J) + b J + c J^2\ .
\eeq
The two corresponding effective potentials are $V(T,\rho)$ and
\beq
\bar{V}(T,\bar{\rho}) = \bar{W}(T,J) - \2\bar{\rho} J\ ,
\eeq
with
\beq
\2\bar{\rho} = {\6\over \6 J}\bar{W}(T,J) = \2\rho + b + 2 c J \ .
\eeq
For stationary points $\rho_0$ and $\bar{\rho}_0$ of the effective potentials,
where
\beq
{\6\over \6\rho}V(T,\rho)\mid_{\rho=\rho_0} = 0\ ,\qquad
{\6\over \6\bar{\rho}}\bar{V}(T,\bar{\rho})
\mid_{\bar{\rho}=\bar{\rho}_0} = 0\ ,
\eeq
one easily verifies
\beq
\bar{\rho}_0 = \rho_0 + 2b\ ,\qquad
\bar{V}(T,\bar{\rho}_0) = V(T,\rho_0)\ .
\eeq
This means that the free energy of the ground state is independent of the two
parameters $b$ and $c$, whereas the expectation value $\rho$ is arbitrary.
However, in the case of more than one stationary point, which is relevant for
first-order phase transitions, the difference $\Delta\rho$ between two
stationary points is independent of $b$ and $c$. This difference is a physical
observable. Other properties of the gauge invariant potential $V(T,\rho)$, such
as the curvature at a local minimum, will in general depend on the parameters
$b$ and $c$.

The functional integral for the free energy may be written as
\bea\label{w}
\exp\left(-\be\Om W(T,J)\right)&\!\!=&\!\!2\pi^2\!\int\ph^3 d\ph\int_{\be}D\PH
D\PH^{\hc} DW_{\mu}\delta\left(\ph - {1\over \beta\Om}\int_\beta dx\,\s\right)
\delta^3\left({1\over \beta\Om}\int_\beta dx\,\vec{\pi}\right) \nn\\
&& \qquad\exp\left(-\int_{\be}dx\left(L + J \PH^{\hc}\PH\right)\right)\ .
\eea
Here the integrand of the ordinary integral over $\ph$ is the exponential of
the well known constraint effective potential \cite{fukuda,oraef},
\bea\label{u}
\exp\left(-\be\Om U(T,\ph;J)\right) &=& \int_{\be}D\PH D\PH^{\hc} DW_{\mu}
\delta\left(\ph - {1\over \beta\Om}\int_\beta dx\,\s\right)
\delta^3\left({1\over \beta\Om}\int_\beta dx\,\vec{\pi}\right) \nn\\
&& \qquad \exp\left(-\int_{\be}dx\left(L + J \PH^{\hc}\PH\right)\right)\ .
\eea
In the infinite volume limit the constraint effective potential $U(T,\ph;J)$
coincides with the effective potential $\tilde{V}(T,\ph;J)$. From eqs.
(\ref{w}),(\ref{u}) one obtains for the free energy,
\beq
\exp\left(-\be\Om W(T,J)\right) = 2\pi^2\int \ph^3d\ph
\exp\left(- \be\Om \tilde{V}(T,\ph;J)\right)\ .
\eeq
In the infinite volume limit, this yields
\beq\label{conn}
W(T,J) = \tilde{V}(T,\ph_{min}(T,J),J)\ ,
\eeq
where $\ph_{min}(T,J)$ is the global minimum of the effective potential
$\tilde{V}(T,\ph;J)$. For arbitrary values of $\ph$ the potential $\tilde{V}$
is gauge dependent. However, its value at the minimum is known to be
gauge independent \cite{kugo}, yielding a gauge independent free energy
$W(T,J)$.

At the critical temperature $T_c$ of
a first-order transition the order parameter $\rho$ and the energy
density $E(T,0)$,
\beq
E(T,J) = W(T,J) - T {\6\over \6 T}W(T,J)\ ,
\eeq
are discontinuous. The jump in the energy density is the latent heat
$\Delta Q$. This
discontinuity in $\rho$ and $E$ requires a free energy $W(T,J)$ with
the following properties: $W(T,0)$ must be continuous but not
differentiable at $T=T_c$, and the same must hold for $W(T_c,J)$
at $J=0$. In the following sections we shall verify these features
based on a perturbative evaluation of the free energy.

\section{Clausius-Clapeyron equation}

For the first-order phase transition from liquid to vapour there exists
a well known relation between the latent heat and the change of the
molar volume, the Clausius-Clapeyron equation \cite{callan}. In the
electroweak phase transition the ``order parameter'' $\PH^{\hc}\PH$
plays the role of the molar volume and a completely analogous relation
can be derived.

The electroweak plasma can exist in two phases, the massive low-temperature
Higgs phase with free energy $W_b(T,J)$ and the massless
high-temperature symmetric phase with free energy $W_s(T,J)$. In the
$J-T$-plane the boundary between the two phases is determined by the
equilibrium condition
\beq
W_s(T,J(T)) = W_b(T,J(T))\ ,
\eeq
which implies
\beq
{\6\over \6 T}\left(W_s - W_b\right) =
- {\6\over \6 J}\left(W_s - W_b\right) {d J(T)\over d T}\ .
\eeq
Using the definitions for latent heat and jump in the order parameter,
\beq
{\6\over \6 T}\left(W_s - W_b\right) = - {1\over T}\D Q \quad,\quad
{\6\over \6 J}\left(W_s - W_b\right) = \2 \D \rho\ ,
\eeq
one obtains
\beq\label{cceq}
\D Q = \2 \D\rho\, T{d J\over d T}\ .
\eeq
This is the Clausius-Clapeyron equation of the electroweak phase transition.

So far, we have only used the continuity of the free energy along the
phase boundary in the case of a first-order transition. We can now employ
the fact that the mass term $\mu + J$ is the only dimensionful parameter
of the SU(2) Higgs model. This implies
\beq
\mu + J(T) = C(g^2, \la) T^2\ ,
\eeq
and therefore
\beq\label{deriv}
T {d J(T)\over d T}\Bigg|_{J=0} = 2 \mu\ .
\eeq
To leading order in the couplings one has (cf. (\ref{1lpot})),
$C(g^2,\la) = - ({3\over 16} g^2 + \2 \la)$. Inserting eq. (\ref{deriv}) into
eq. (\ref{cceq}) we finally obtain\footnote{The same result has been
derived in a recent paper by Farakos et al. \cite{fkrs} based
on renormalization group equations for the effective three-dimensional
theory.}
\beq
\D Q = - \2 m_H^2 \D \rho (1 + {\cal O}(g^2, \la))\ .
\label{cc}\eeq
The higher order corrections are due to the difference between the mass
parameter $\sqrt{-2\mu}$ and the physical Higgs mass. In the following sections
this relation will provide a very useful check on our results.

\section{Free energy in perturbation theory}

Near the ground state, $J = 0$,
the free energy $W(T,J)$ can be evaluated as power series in the couplings
$g$ and $\la$ by means of resummed perturbation theory\footnote{For a
detailed discussion and references, see \cite{bfhw}.}. Here thermal
corrections are added to the tree-level masses of the scalar fields
and the longitudinal component of the vector boson field,
\beq \label{CT}
\delta S_{\be} = \be \int d^3x \left(\2 \a_{01} T^2 (\s^2 + \pi^2)
+ \2 \a_1 T^2 W_L^2\right)\ .
\eeq
The sum of tree-level masses and thermal corrections then enters the boson
propagators in loop diagrams, and $\de S_{\be}^c = -\de S_{\be}$ is treated as
counter term. In eq. (\ref{CT}) the fields $\s$ and $\pi$ do not depend on the
imaginary time $\tau$. Following Arnold and Espinosa \cite{arnold}, we only
resum the static modes of scalar and vector fields. Clearly, this is sufficient
to avoid infrared singular contributions from these fields since non-static
modes have thermal masses ${\cal O}(T)$. For the usual resummation with
one-loop thermal counter terms $\sim T^2$ the resummation of static modes is
known to be equivalent to the resummation of all modes. However, in the
following we shall employ counter terms which depend on the scalar background
field. It turns out that in this case the resummation of static modes is
preferred, since only for this resummation the loop expansion is an expansion
in the couplings $g$ and $\la$.

To leading order in the couplings, one obtains for the parameters
in eq. (\ref{CT}) from
one-loop self energy corrections \cite{bfhw},
\beq
\a_{01} = {3\over 16}g^2 + \2 \la \quad , \qquad \a_1 = {5\over 6}g^2\ .
\eeq
The masses of the boson propagators are obtained from eqs. (\ref{ST}) and
(\ref{CT}) by shifting the Higgs field $\s$ by the average field
$\ph$ (cf. (\ref{w}),(\ref{u})). This yields $m_L$, $m_T$,
$m_{\s}$ and $m_{\pi}$ for longitudinal and transverse part of the
vector field, the Higgs field and the Goldstone boson field,
\bea\label{masses2}
m_L^2 = \a_1 T^2 + {g^2\over 4}\ph^2\quad ,\qquad\hspace{1.4cm}\mbox{}
&& m_T^2  =  {g^2\over 4}\ph^2\ ,\label{mlt}\\
m_{\s}^2 = \a_{01} T^2 + \mu + J + 3 \la \ph^2\quad ,\qquad
&& m_{\pi}^2 = \a_{01} T^2 + \mu + J +\la \ph^2\  \label{ms}\ .
\eea
The scalar masses agree with derivatives of the effective potential
$\tilde{V}$,
\beq\label{mex}
m_{\s}^2 = {\6^2\over \6\ph^2}\tilde{V}(T,\ph;J)\quad ,\qquad
m_{\pi}^2 = {1\over \ph}{\6\over \6\ph}\tilde{V}(T,\ph;J)\ ,
\eeq
up to terms of higher orders in the couplings $g$ and $\la$.

It has been shown that the resummed loop-expansion for the
free energy is a systematic expansion in the couplings $g$ and $\la$
\cite{bfhw}. So far, however, several aspects of this
resummation have remained
unsatisfactory. First, higher order corrections are very important for
the scalar masses\footnote{We thank P.~Arnold for emphasizing this problem.}.
In particular, in the Higgs phase the Goldstone
boson mass $m_{\pi}$, as given by eq. (\ref{mex}), vanishes at the
minimum. Hence, the expression to leading order in the couplings,
given in eq. (\ref{ms}), is cancelled by higher order corrections.
In the following we shall therefore modify the loop expansion in the
Higgs phase. We replace eq. (\ref{ms}) by
\beq\label{msb}
m_{\s}^2 = 2\la \ph^2 \quad ,\qquad m_{\pi}^2 = 0\ ,
\eeq
and we treat
\beq\label{ctb}
\de S_{\be}^b = \be\int d^3x \2 \left(\a_{01}T^2 +\mu + J + \la \ph^2
\right)(\s^2 + \pi^2)\ ,
\eeq
as counter term, where $\s$ and $\pi$ again represent static modes.
Note, that in the sum
$\de S_{\be}^c + \de S_{\be}^b$ the temperature dependent terms cancel. Hence,
we perform no thermal resummation for scalar masses in the Higgs
phase.

In the symmetric phase, $\ph = 0$, and at temperatures close to the
barrier temperature $T_b = \sqrt{\mu/\a_{01}}$,
the one-loop finite-temperature scalar masses
are small, and higher order corrections are important for the Higgs boson
and the Goldstone boson mass. Here, we will replace eqs. (\ref{ms})
by self-consistently determined scalar masses, which are defined by
\beq\label{selfc}
m_{\s}^2 = m_{\pi}^2 = {1\over \ph}{\6\over \6\ph}\tilde{V}(T,0;J)\ .
\eeq
For given vector boson masses $m_L$ and $m_T$, this is a gap equation
for the scalar masses, which can be solved at each order of the loop
expansion.

The self-consistent scalar mass at one-loop is easily obtained from the
one-loop effective potential which reads, for arbitrary resummations
of the static modes,
\bea\label{1lpot}
\tilde{V}(T,\ph;J) &=& \frac{T^2}{6}J+\2(\a_{01} T^2 + \mu +J)\ph^2 + \4\la
\ph^4 \nn\\ && - {T\over 12\pi}\left(3m_L^3 + 6m_T^3 + m_{\s}^3 + 3m_{\pi}^3
\right) + {\cal O}(g^4,\la^2)\ .
\eea
{}From eqs. (\ref{mlt}), (\ref{selfc}) and (\ref{1lpot}) one obtains in the
symmetric phase
\beq\label{mss}
m_{\s}^2 = m_{\pi}^2 = \a_0 T^2 + \mu + J + {
\cal O}\left({g^2m_T},
{\la m_{\s}}\right)\ ,
\eeq
where
\beq
\a_0 = {3\over 16}g^2 +\2\la - {3\over 16\pi}\sqrt{{5\over 6}}g^3\ .
\eeq
The corresponding counter term for perturbation theory in the symmetric
phase, to be inserted in one-loop graphs in a two-loop calculation
for $W(T,J)$, reads
\beq\label{cts}
\de S_{\be}^s = - \be \int d^3x \left(\2\a_0 T^2 (\s^2 + \pi^2)
+ \2 \a_1 T^2 W_L^2\right) \ .
\eeq

Having specified the vector boson and scalar masses in the Higgs phase and the
symmetric phase, as given in eqs. (\ref{mlt}), (\ref{msb}) and
(\ref{mss}), the free energy can be calculated from the effective potential
$\tilde{V}$, using the relation (\ref{conn}).
The effective potential has two local minima,
\beq
\ph_{min,1}(T,J) \equiv \ph_s = 0\ ,\quad \ph_{min,2}(T,J) \equiv \ph_b > 0\ ,
\eeq
which correspond to the symmetric and the Higgs phase, respectively.
{}From eq. (\ref{1lpot}) one obtains for the free energy in both cases
\bea
W_s(T,J) &=& \frac{T^2}{6}J
             -{T\over 3\pi}(\a_0 T^2 + \mu + J)^{3/2}\ ,\nn\label{ws1}\\
W_b(T,J) &=& \tilde{V}(T,\ph_b(T,J);J) \\
         &=& \frac{T^2}{6}J+\2(\a_{01}T^2 + \mu +J)\ph_b^2 + \4\la\ph_b^4 \nn\\
         &&\!\!\!-{T\over 12\pi}\left(6\left(\4 g^2 \ph_b^2\right)^{3/2}
               \!\!\!+ \left(2\la \ph_b^2\right)^{3/2}
               \!\!+ 3\left(\a_1 T^2 + \4 g^2\ph_b^2\right)^{3/2}
               \!\!\!\!-3\a_1^{3/2} T^3 \right)\!\!.\label{ws2}
\eea
For convenience, we have subtracted the terms independent of $J$ from
$W_s$ and $W_b$.

The free energy of the ground state is given by
\beq
W(T,0) = \min\{W_s(T,0), W_b(T,0)\}\ .
\eeq
It is a concave function, shown in fig. 1 for some choice of $g$ and $\la$. The
derivative of $W(T,0)$ has a jump at the critical temperature $T_c$, which is
characteristic for a first-order phase transition. The corresponding latent
heat is given by
\beq
\Delta Q = T{d\over dT}\left(W_b(T,0)-W_s(T,0)\right)\mid_{T=T_c}
  = -\mu \left(\ph_b^2 + {T_c\over \pi}\left(\a_0 T_c^2 + \mu\right)^{1/2}
\right)\ ,
\eeq
which follows immediately from the Clausius-Clapeyron equation or, with some
more work, from eqs. (\ref{ws1}),(\ref{ws2}).
We can also consider $W(T_c,J)$, i.e., the dependence of the free energy
on the external source at the critical temperature $T_c$. This function,
shown in fig. 2, is also concave and similar to the function plotted
in fig. 1. In this plot the huge linear term $T^2J/6$, which has only
the effect to shift the field square expectation value by $T^2/3$, has been
discarded in both phases. As discussed in sect. 2,
the jump in the derivative at $J=0$ yields the jump in the
``order parameter'' $\rho$ at the first-order phase transition.
One easily verifies that the Clausius-Clapeyron equation (\ref{cc})
is satisfied.

{}From the free energy $W(T,J)$ one can obtain the gauge invariant
effective potential $V(T,\rho)$ by means of a Legendre transformation.
Since the derivative of $W(T,J)$ is not continuous everywhere, one
has to use the definition\footnote{For a discussion and references,
see \cite{oraef}.},
\beq
V(T,\rho) = \sup_J\{W(T,J) - \2 \rho J\}\ .
\eeq
This yields the convex, non-analytic function, plotted in fig. 3 as
full line. One may also compute the ordinary Legendre transform
$V_s(T,\rho)$ and $V_b(T,\rho)$ of $W_s(T,J)$ and $W_b(T,J)$, respectively.
Neglecting constant terms this yields
\bea
V_s(T,\rho) &=& \2(\a_0 T^2 + \mu)\rho' - {\pi^2\over 6 T^2}{\rho'}^3\ ,\nn\\
V_b(T,\rho) &=& \tilde{V}(T,\sqrt{\rho'},0)\ ,
\eea
where we have used
\beq
\2\rho' \equiv \2\left(\rho-\frac{T^2}{3}\right)={\6 W_b(T,J)\over \6 J}-
\frac{T^2}{6}= \2 \ph_b^2 + {\6 \tilde{V}\over \6 \ph}\Bigg|_{\ph_b} {\6 \ph_b
\over \6 J} = \2 \ph_b^2\ .
\eeq
$V_s$ and $V_b$ are also shown in fig. 3. In the region outside of the two
local minima, $V_s$ and $V_b$, respectively, coincide with the convex effective
potential $V(T,\rho)$. Between the two local minima, $V_s$ and $V_b$ represent
two analytical continuations of $V(T,\rho)$, which meet at the ``matching
point'' $\rho_M = T^2/3$. At this point, marked by a cross in the plot, the
first derivatives of both curves coincide.

The non-convex ``effective potential'' obtained by combining $V_s$ and $V_b$ on
both sides of the ``matching point'' is almost identical with the
gauge invariant ``effective potential'' obtained in \cite{bfh}. If the root
appearing in $V_b(T,\rho)$ is expanded to order $\rho^2/T^4$, which corresponds
to the reduction to the three-dimensional theory considered in \cite{bfh}, both
``effective potentials'' are identical. The generation of a barrier between two
local minima as analytic continuation from a convex effective potential is
reminiscent of the treatment of first-order phase transitions in condensed
matter physics \cite{langer}. However, the precise physical meaning of the
resulting non-convex ``effective potential'' still remains to be understood.

In \cite{bfh} it was pointed out that the usual Landau-gauge effective
potential and the gauge invariant effective potential lead to different
predictions for observables like latent heat, critical temperature etc. Our
derivation of the gauge invariant effective potential in this section
demonstrates that these differences are a consequence of different choices of
the resummation procedure. Based on the arguments given above, the asymmetric
resummation which treats symmetric phase and Higgs phase differently, appears
better justified.

The results of \cite{bfh} were obtained by performing an expansion around the
tree-level minima in the Higgs phase and in the symmetric phase. This approach
has been considered at two-loop level \cite{fkrs,laine1}, and some problems
have been discussed by Laine \cite{laine1}. In this section we have reproduced
the results of \cite{bfh} by expanding around local minima of the effective
potential $\tilde{V}$ which includes quantum corrections. As we shall
demonstrate in the following section, this procedure can be extended to two
loops. At this level the fundamental infrared problem of the symmetric phase
will also become apparent, and we shall discuss under which conditions the loop
expansion can yield a good approximation of thermodynamic quantities relevant
for the electroweak phase transition.

\section{Two-loop results}\label{tl}

In this section we shall extend the asymmetric resummation to two loops
in order to examine the convergence of the perturbative expansion.
In recent years several two-loop calculations have already been carried out.
In \cite{arnold} all two-loop contributions involving only the gauge coupling
were evaluated, which yield the effective potential $\tilde{V}$ to order
$g^4, \la$. In this calculation scalar masses were set equal to zero.
A complete two-loop calculation of $\tilde{V}$ to order $g^4, \la^2$,
including scalar loops, has been carried out in \cite{hebecker}, where also the
full standard model has been considered. Here the
masses (\ref{masses2}),(\ref{ms}), with $J = 0$,
were used in the symmetric phase
and the Higgs phase. The same result has also been obtained by
using as intermediate step the effective three-dimensional theory, which
is obtained by integrating out modes with non-zero Matsubara frequencies
\cite{farakos}. Furthermore, the effective potential of the
three-dimensional theory
has been evaluated in general covariant \cite{laine} and 't Hooft
background gauges \cite{schmidt}.

In all these calculations the resummation of scalar masses
has been performed
in the same way in the symmetric and the Higgs phase. However, on
physical grounds, as explained in the previous section, an asymmetric
treatment of the two phases appears more appropriate. In the
Higgs phase scalar masses are given by eq. (\ref{msb}) and the counter term
is defined by the sum $\de S_{\be}^c + \de S_{\be}^b$, as discussed in sect. 4.
In the symmetric phase the scalar masses are
self-consistently determined from eq. (\ref{selfc}),
\beq
m_{\s}^2 = m_{\pi}^2 = {1\over \ph}{\6\over \6\ph}\tilde{V}(T,0;J)\ ,\nn
\eeq
where the potential $\tilde{V}$ is calculated with the counter term
$\de S^s_{\be}$ given in eq. (\ref{cts}). Note, that in the two-loop
calculation the counter term to be inserted in the one-loop graph
is ${\cal O}(g^3)$, whereas the scalar mass determined from eq. (\ref{selfc})
is of higher order in $g$.

The two-loop potential $\tilde{V}$ for arbitrary masses $m_{\s}$ and
$m_{\pi}$ can be extracted from \cite{arnold,hebecker}. We have listed the
individual contributions corresponding to the graphs of fig. 9
in the appendix, omitting terms independent of $\ph$ and terms which cancel
in the sum. The two-loop potential contains
terms linear in $m_{\s}$ and $m_{\pi}$. In the symmetric phase, where the
counter term is given by (\ref{cts}), these terms cancel in the sum. In
the Higgs phase they contribute to the potential.

In the symmetric phase the scalar masses are determined self-consistently
by eq. (\ref{selfc}). With $m_T = g\ph/2$, the two-loop potential yields
a contribution which diverges logarithmically at $\ph \approx 0$
(cf. (\ref{2dsing})),
\beq\label{ldiv}
m^2_{\s} \approx - {33 g^4\over 128\pi^2}T^2 \ln{\be m_T}\ .
\eeq
Following \cite{bfhw} we regularize this divergence by means of a
``magnetic mass'' term. In eq. (\ref{ldiv}) we substitute
$m_T^2 = g^2\ph^2/4 + \gamma^2g^4 T^2/(9\pi^2)$. In the following we
shall use $\gamma =1$, which follows from one-loop gap equations
\cite{bfhw,espinosa}. We have checked that the results of our
numerical analysis change only insignificantly if we vary the parameter
$\gamma$ between 0.3 and 3.0. The deviation of the most sensitive quantity
$\Delta Q/T_c^4$ from the plotted $\gamma=1$ result is invisible for small
Higgs mass and increases up to $8\%$ at $m_H=$ 70 GeV. Eq. (\ref{selfc}) for
the scalar masses can be solved iteratively. In the first step one inserts
in the two-loop potential $\tilde{V}$ the one-loop scalar masses
(\ref{masses2}). The mass $m_{\s}$ obtained in this first step of
iteration is already a very good approximation to the exact solution
in the range of Higgs masses which we shall consider.

Given the two-loop potential in the symmetric phase and the Higgs phase,
we can numerically determine the critical temperature $T_c$, where
the two potentials at their respective local minima are degenerate.
Differentiation with respect to temperature and the external source $J$
then yields latent heat and jump in the order parameter $\rho = 2\PH^{\hc}\PH$.
The results are shown in figs. 4 - 6, assuming standard model values
$m_W=$ 80.22 GeV and $v=$ 251.78 GeV at zero temperature.

In fig. 4 the critical temperature in units of the Higgs mass is plotted as
function of the Higgs mass in the range 30 GeV$<m_H<$80 GeV. Below 30 GeV the
high-temperature expansion is unreliable, and above 80 GeV the convergence of
the loop expansion deteriorates rapidly. The two-loop results are compared with
one-loop results for the ``old resummation'' \cite{hebecker} and for the ``new
resummation'' described above. In the ``new resummation'' the one-loop result
is lowered whereas the two-loop result is increased, improving the convergence
of the loop expansion considerably. The relative change of an observable from
one-loop to two-loop may be characterized by $\de = 2 |O_1 - O_2|/(O_1 + O_2)$.
In the ``new resummation'' $\de \sim 0.04$ in the whole range of Higgs masses
considered.

The jump in the order parameter at the critical temperature $T_c$ is shown in
fig. 5. In the case of ``old resummation'' $\varphi_c$ corresponds to the
position of the second minimum, for ``new resummation'' it is $\sqrt{\Delta
\rho}$. Again the ``new resummation'' procedure improves the convergence
significantly. The relative error increases from $\de \sim 0.01$ at $m_H = 40$
GeV to $\de \sim 0.2$ at $m_H = 70$ GeV.

In fig. 6 the latent heat $\Delta Q$ in units of the critical temperature is
plotted as function of the Higgs mass. This is a measure of the strength of the
first-order phase transition. Like the jump in the order parameter it decreases
with increasing Higgs mass. Based on the Clausius-Clapeyron equation, which is
satisfied exactly, we expect $\Delta Q \sim \Delta \rho$. Hence, the
convergence should be worse than for the order parameter. This is indeed the
case. The relative error increases from $\de \sim 0.1$ at $m_H = 40$ GeV to
$\de \sim 0.4$ at $m_H = 70$ GeV.

It appears satisfactory that the new, asymmetric resummation procedure leads
to an improved convergence of the perturbative expansion. The ``old
resummation'' is based on a systematic expansion of the free energy in powers
of the couplings $g$ and $\la$. Therefore in the symmetric phase, where the
scalar masses are positive, the differences can only correspond to
contributions of higher order. This ambiguity in the resummation procedure is
analogous to the well known dependence of results in perturbative QCD on the
renormalization scheme. However, setting the Goldstone mass to zero in the
Higgs phase does not correspond to a simple rearrangement of the series, since
terms non-analytic in $m_\pi$ are present. The correct treatment of the Higgs
phase is the main qualitative advantage of the new resummation.

\section{Comparison with lattice simulations}

In the previous sections we have presented a quantitative description of the
electroweak phase transition in the perturbative approach. Using different
resummations in the Higgs phase and in the symmetric phase, we concluded that
for Higgs boson masses below $\sim 70$ GeV the perturbative approach converges,
and that this description is therefore self-consistent. However, for Higgs
masses above the present experimental lower bound, $m_H>63$ GeV, the loop
expansion becomes unreliable. In this relevant range of the parameter space the
electroweak phase transition can only be understood by means of
non-perturbative methods. Lattice Monte Carlo simulations provide a well
defined and systematic approach to study this problem. For all Higgs masses,
non-perturbative effects may be important in the symmetric phase. By comparing
data from lattice simulations for the SU(2) Higgs model with the perturbative
results one can hope to identify non-perturbative features and to achieve a
better understanding of the electroweak phase transition. Therefore, in this
section we shall present a detailed comparison between data of the recent large
scale four-dimensional lattice works \cite{fodor,montvay} and published results
of the ``old resummation'' \cite{hebecker}. For the considered Higgs masses
they differ little from the ``new resummation'' results.

Monte Carlo simulations for Higgs boson masses near or above the W-boson mass
are technically difficult, thus in \cite{fodor,montvay} the Higgs masses $m_H
\approx 18$ GeV and $m_H\approx 49$ GeV have been studied. Since in this
parameter range the perturbative expansion converges rather well, one may
expect agreement between perturbative and non-perturbative results with
comparable accuracy.

The $g^3,\la^{3/2}$-potential of \cite{hebecker} involves a high-temperature
expansion up to order $(m/T)^3$, which is unsatisfactory for $m_H\approx 18$
GeV. Thus, we have included all one-loop contributions of order $(m/T)^4$ in
our present $g^3,\la^{3/2}$-potential. Note, that the numerical evaluation of
the one-loop temperature integrals gives a result which agrees with the above
approximation up to a few percent.

For small Higgs boson masses the renormalization scheme dependence is
non-negligible. Therefore, instead of the $\overline{\mbox{MS}}$-scheme with
$\bar{\mu}=T$, we shall use the scheme suggested by Arnold and Espinosa
\cite{arnold}, which includes the most important zero-temperature
renormalization effects. In this scheme the correction to the
$\overline{\mbox{MS}}$-potential, used for both the one- and the two-loop
results, reads
\beq
\delta V={\varphi^2 \over 2} \left( \delta\mu + {1 \over 2\beta^2}
\delta\lambda\right) + {\delta\lambda \over 4} \varphi^4,
\eeq
where
\beq
\delta \mu = {9g^4v^2 \over 256 \pi^2},\ \ \ \ \ \
\delta\lambda=-{9g^4\over 256\pi^2}\left(\ln\frac{m_W^2}{\bar{\mu}^2}+{2\over
3} \right).
\eeq
Here $v$ is the zero-temperature vacuum expectation value and $m_W$ is the
W-boson mass at $T=0$.

In \cite{montvay} several observables have been determined, including
renormalized masses at zero temperature ($m_H$, $m_W$), critical temperature
($T_c$), jump in the order parameter ($\varphi_c$), latent heat ($\Delta Q$)
and surface tension ($\sigma$). As usual, the dimensionful quantities have been
normalized by the proper power of the critical temperature. The simulations
have been performed on $L_t=2$ and $L_t=3$ lattices ($L_t$ is the temporal
extension of the finite-temperature asymmetric lattice). The $L_t=3$ results
should be closer to the continuum values, and we therefore compare these data
with the perturbative results. An exception is the surface tension for which
only $L_t=2$ data exist.

In the previous sections we have only evaluated $T_c$, $\varphi_c$ and $\Delta
Q$, which follow from the free energy of a homogeneous phase. The surface
tension is more complicated to calculate, since it involves the boundary
between two phases. In perturbation theory this is related to the effective
potential $\tilde{V}$ in the region between the two local minima. For
completeness, we include the surface tension in the comparison, although so far
no satisfactory treatment has been achieved in perturbation theory for Higgs
masses above $\sim 40$ GeV.

\renewcommand{\arraystretch}{1.5}\label{CC}
\begin{table}[b]
\begin{center}
\begin{tabular}{|l||l|l|}
\hline
& $m_H\approx 18$ GeV & $m_H\approx 49$ GeV  \\ \hline\hline
$\Delta Q$ from Clausius-Clapeyron eq. & .0236(14) & .00171(15) \\ \hline
direct lattice result for $\Delta Q$  & .0194(15) & .00151(12) \\ \hline
\end{tabular}
\end{center}
\caption[]{\it Comparison of the latent heat in lattice units obtained by using
the Clausius-Clapeyron equation and lattice Monte Carlo simulations. The data
are from ref. \cite{montvay}.}
\end{table}

The statistical errors of these observables are normally determined by
comparing statistically independent samples. The systematic errors can be
estimated by the difference between the $L_t=2$ and the $L_t=3$ data. As fig.
15 of \cite{montvay} suggests, the true systematic error for $T_c/m_H$ may be
larger than this naive estimate. Thus, for $T_c/m_H$ we have doubled the above
error. For the surface tension, where only $L_t=2$ data exist, the
systematic error has been estimated as twice the statistical one. A correct
comparison has to include errors on the parameters used in the perturbative
calculation. These uncertainties are connected with the fact that neither the
Higgs boson mass nor the gauge coupling has been determined exactly. Therefore,
the perturbative prediction for an observable is not one definite value but
rather an interval, given by the uncertainties of $m_H$ and $g$. In this
analysis only the statistical errors in the determination of the Higgs boson
mass and the gauge coupling have been included. We have also neglected
corrections due to the different renormalization conditions used for the gauge
coupling $g$ and the order parameter $\PH^{\hc}\PH$ on the lattice and in the
continuum. These corrections are expected to be of relative order $g^2$.

Let us first check the validity of the Clausius-Clapeyron equation for the
lattice Monte Carlo simulations. The first line of table 1 contains $\Delta Q$
obtained by eq. (\ref{cc}) from lattice results for $\Delta\rho$ and $m_H^2$.
The second line contains the latent heat determined directly by lattice
simulations. As usual, the numbers in the parentheses denote the statistical
errors. The values for $m_H \approx 49$ GeV agree within one standard
deviation. For $m_H \approx 18$ GeV, however, there is a small discrepancy.

The comparison  for $\varphi_c/T_c$, $\Delta Q/T_c^4$, $\sigma/T_c^3$ and $T_c
/m_H$ is shown in fig. 7 ($m_H \approx 18$ GeV) and in fig. 8 ($m_H \approx 49$
GeV). The dots with error bars represent the perturbative result at one-loop
($g^3$) and at two-loop ($g^4$) level. For each quantity the dashed lines show
the region allowed by the statistical error, whereas the dotted lines include
the systematic error as well. Note, that the values of $g$ and $m_H/m_W$ are
those of \cite{montvay}. We emphasize that, both in the perturbative
calculation and in the lattice simulations, the surface tension is the most
problematic quantity.

For $m_H \approx 18$ GeV (cf. fig. 7) both the one-loop and the two-loop
results are in good agreement with the lattice data. For $m_H \approx 49$ GeV
(cf. fig. 8) the two-loop results agree definitely better with the Monte Carlo
data, except for the surface tension. The two plots may be interpreted in the
following way. For small Higgs boson masses the perturbative approach is in
very good shape, already the one-loop approximation gives a reliable result. As
$m_H$ grows, the higher order contributions become more and more important, yet
a two-loop calculation is still satisfactory for $m_H \approx 49$ GeV. In this
range of parameters the non-perturbative features of the symmetric phase are
not important enough to destroy the perturbative picture.

\section{Conclusions}
In the present paper a quantitative description of the electroweak phase
transition based on a gauge invariantly coupled source term has been attempted.
Our main result is that for Higgs masses below $\sim 70$ GeV thermodynamic
observables can be evaluated with reasonable accuracy in perturbation
theory. Above $m_H\sim 70$ GeV the perturbative expansion breaks down,
which is in agreement with previous estimates \cite{bfhw}.

The main technical achievement is the realization of different resummation
procedures in the symmetric phase and in the Higgs phase, which are expected
to improve the convergence of perturbation theory. Such an improvement has
been explicitly verified for the available terms up to two-loop order.
Due to infrared divergencies of the non-abelian theory a cutoff is needed in
the symmetric phase at two-loop order. We emphasize that due to the small
cutoff dependence at not too large Higgs masses perturbation theory may work
with some accuracy in the symmetric phase. The size of  non-perturbative
effects can only be determined by comparing perturbative results with fully
non-perturbative lattice simulations.

We have carried out such a comparison, based on results of recent simulations
on large lattices \cite{montvay}. The good quantitative agreement found for
Higgs mass values $m_H\approx 18$ GeV and $m_H\approx 49$ GeV is interpreted as
evidence for the correctness of the present understanding of the electroweak
phase transition. Non-perturbative effects present in the symmetric phase
are neglected by perturbation theory, but they should contribute to the lattice
results. We conclude that these effects can not be of major importance at small
Higgs mass, since otherwise no quantitative agreement with lattice data
could be observed.

Applying the Clausius-Clapeyron equation to the electroweak phase transition a
simple relation between latent heat and jump of the order parameter has been
derived. Being in good agreement with perturbative as well as with lattice
data, it strengthens confidence in the correctness of the treatment of the
phase transition.

The above arguments support the conclusion that our understanding of the
electroweak phase transition has reached a quantitative level for Higgs masses
up to $\sim 70$ GeV. A strong decrease of the strength of the first-order
transition with increasing Higgs mass is observed. However, a complete
understanding of the process of symmetry restoration for large Higgs masses is
still lacking. Other important questions include the description of metastable
and unstable states, relevant for the dynamics of the transition.

\section{Appendix}
In this appendix, for the convenience of the reader, all the contributions to
the effective potential from one and two-loop graphs (see fig. 9) are listed in
$\overline{\mbox{MS}}$-scheme. This calculation has already been performed in
\cite{arnold} up to order $g^4$ and extended in \cite{hebecker,farakos} to the
order $g^4,\la^2$. The different contributions are presented according to the
counterterm method of resummation, applied to the zero modes only
\cite{arnold}. The scalar resummation is left unspecified, since it is
different in the symmetric phase and in the Higgs phase following the approach
of section \ref{tl}. It is characterized by the values of the scalar masses
$\ms$ and $\mg$, and the scalar counter term $C_s$, present in $V_a$.

Tree level potential and one-loop corrections are collected in $V_1$ while the
two-loop contributions $V_a$ ... $V_k$ are labelled in correspondence with the
diagrams of fig. 9. Linear mass terms, poles in $\epsilon$ and terms
proportional to $\iota_\epsilon$ (see ref.~\cite{arnold}), which cancel
systematically in the final result, are not shown and the limit $\epsilon\to 0$
has already been performed. This is the reason for the vanishing of $V_b$,
which would only contribute to the well known cancellation of linear mass terms
of order $g^3$.

\begin{eqnarray}
  V_1&\!\!\!=&\!\!\!\frac{\vi^2}{2}\left[\mu+\frac{1}{\beta^2}\left(\frac{1}
  {2}\la+\frac{3}{16}g^2\right)\right]+\frac{\la}{4}\vi^4-\frac{1}{12\pi\beta}
  \left[{\ms}^{3}+3\,{\mg}^{3}+6\,{ m}^{3}+3\,{ m_{L}}^{3}\right]\\
&&\nonumber\\
  &&\!\!\!\!\!\!\!\!\!\!\!\!-\frac{1}{64\pi^2}\left\{\left(12\la^2\vi^4+12\la
  \mu\vi^2+9m^4\right)\left(\ln\bar{\mu}^2\beta^2-c_1+\frac{3}{2}\right)-6m^4
  -\frac{59}{12\beta^2}g^2m^2\right\}\nonumber
\\
&&\nonumber\\
&&\nonumber\\
  V_a&\!\!\!=&\!\!\!\frac{1}{8\pi\beta}(\ms+3\mg)\, C_s\quad,\quad V_b=0
\\
&&\nonumber\\
&&\nonumber\\
  V_c&\!\!\!=&\!\!\!  \!\frac{\la}{\beta^2}\Bigg[\!-\!\frac{\ms+\!3\mg}{16\pi
  \beta}\!+\!\frac{1}{64\pi^2}\!\left\{3\ms^2+6\ms\mg+15\mg^2-6\la\vi^2\!
  \left(\ln\bar{\mu}^2\beta^2-c_1+\frac{3}{2}\right)\!\right\}\!\!\Bigg]
\\
&&\nonumber\\
&&\nonumber\\
  V_d&\!\!\!=&\!\!\!\frac{3g^2}{128\beta^2}\Bigg[-\frac{\ms+3\mg}{\pi\beta}\\
&&\nonumber\\
  &&\!\!\!\!\!\!\!\!\!\!\!\!+\frac{1}{\pi^2}\left\{(\ms+3\mg)(2m+m_L)-
  \frac{1}{4}\left(4m^2+6\la\vi^2\right)\left(\ln\bar{\mu}^2\beta^2-c_1+
  \frac{5}{6}\right)\right\}\Bigg]\nonumber
\\
&&\nonumber\\
&&\nonumber\\
  V_e&\!\!\!=&\!\!\!\frac{3g^2}{256\pi^2\beta^2}\Bigg[32m_Lm-9m^2\left(\ln
  \bar{\mu}^2\beta^2-c_1-\frac{125}{54}\right)\Bigg]
\\
&&\nonumber\\
&&\nonumber\\
  V_f&\!\!\!=&\!\!\!\frac{3g^2}{128\pi^2\beta^2}\;\Bigg[2m^2\left(\ln{\frac{
  \beta}{3}}-\frac{1}{12}\ln{\bar{\mu}^2\beta^2}-\frac{1}{6}c_1+\frac{1}{4}c_2+
  \frac{1}{4}\right)+\mg\left(\ms+\mg\right)\\
&&\nonumber\\
  &&\!\!\!\!\!\!\!\!\!\!\!\!+\!\frac{1}{2}\left(\ms^2\!+3\mg^2\right)\left(\!
  -4\ln{\frac{\beta}{3}}\!+\ln{\bar{\mu}^2\beta^2}\!-c_2\right)\!-\frac{1}{m}
  (\ms\!-\mg)^2(\ms+\mg)\!-m(\ms+3\mg)\nonumber\\
&&\nonumber\\
  &&\!\!\!\!\!\!\!\!\!\!\!\!-\frac{1}{m^2}\left(\ms^2-\mg^2\right)^2\ln(\ms+
  \mg)+\left(m^2-4\mg^2\right)\ln(2\mg+m)\nonumber\\
&&\nonumber\\
  &&\!\!\!\!\!\!\!\!\!\!\!\!+\frac{1}{m^2}\left\{m^4-2\left(\ms^2+\mg^2\right)
  m^2+\left(\ms^2-\mg^2\right)^2\right\}\ln{(\ms+\mg+m)}\Bigg]\nonumber
\\
&&\nonumber\\
&&\nonumber\\
  V_{gh}&\!\!\!=&\!\!\!\frac{g^2}{16\pi^2\beta^2}\Bigg[\frac{m^2}{8}\left(31
  \ln\bar{\mu}^2\beta^2-66\ln m+39\ln 3-\frac{11}{2}c_1-\frac{51}{2}c_2-
  \frac{145}{4}-102\ln\beta\right)\nonumber\\
&&\nonumber\\
  &&\!\!\!\!\!\!\!\!\!\!\!\!-3m_Lm+\frac{3}{2}m_L^2+\left(\frac{3}{2}m^2-6m_L^2
  \right)\ln(2m_L+m)\Bigg] \label{2dsing}
\\
&&\nonumber\\
&&\nonumber\\
  V_j&\!\!\!=&\!\!\!-\frac{3\la^2\vi^2}{32\pi^2\beta^2}\left[\ln
  \frac{9\bar{\mu}^2}{\beta^2}-c_2-2\ln\{\ms(\ms+2\mg)\}\right]
\\
&&\nonumber\\
&&\nonumber\\
V_k&\!\!\!=&\!\!\!\frac{3g^2}{64\pi^2\beta^2}\Bigg[m^2\left(\frac{5}{4}\ln
  \frac{\beta}{9\bar{\mu}}+\frac{5}{8}c_2+\ln(2m_L+\ms)\right)+\frac{\ms^4}
  {4m^2}\ln\ms+\frac{1}{2}\ms m\\
&&\nonumber\\
  &&\!\!\!\!\!\!\!\!\!\!\!\!-\frac{1}{2m^2}(m^2-\ms^2)^2\ln(m+\ms)+\left(2m^2
  -\ms^2+\frac{\ms^4}{4m^2}\right)\ln(2m+\ms)+\frac{1}{4}\ms^2\Bigg]\, .
  \nonumber
\end{eqnarray}
Here evaluating the scalar one- and two-loop temperature integrals the
constants
    \begin{equation} c_1\approx 5.4076\quad\mbox{and}\quad c_2\approx 3.3025
    \end{equation}
have been introduced following \cite{jackiw} and \cite{parwani,zhai},
respectively.

\newpage
\noindent
\\
{\bf\large Figure captions}\\
\\
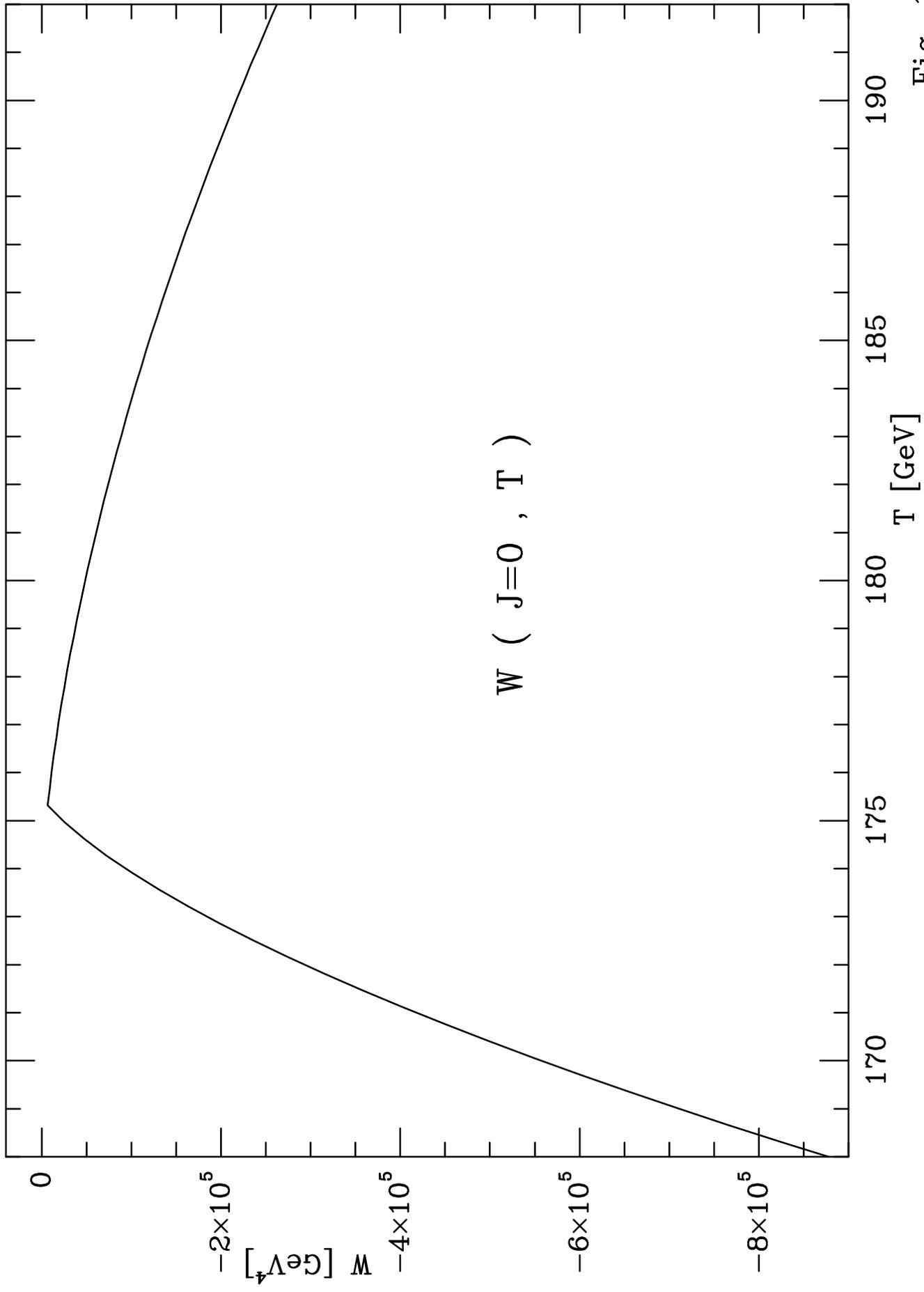
{\bf Fig.1} Free energy at zero source term as a function of the temperature
\newline($m_H$ = 70 GeV).\\
\\
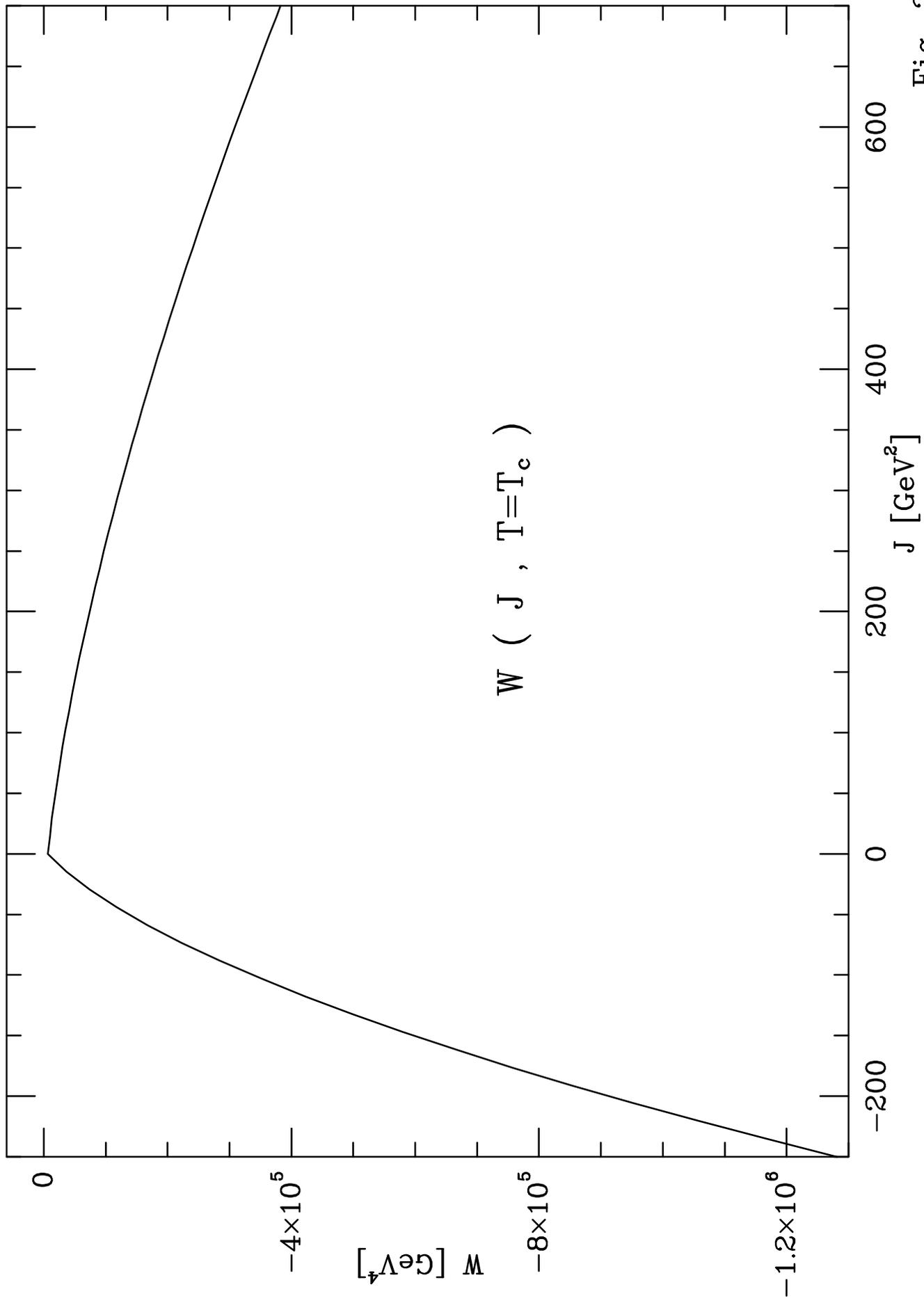
{\bf Fig.2} Free energy at the critical temperature as a function of the source
$J$ \newline($m_H$ = 70 GeV).\\
\\
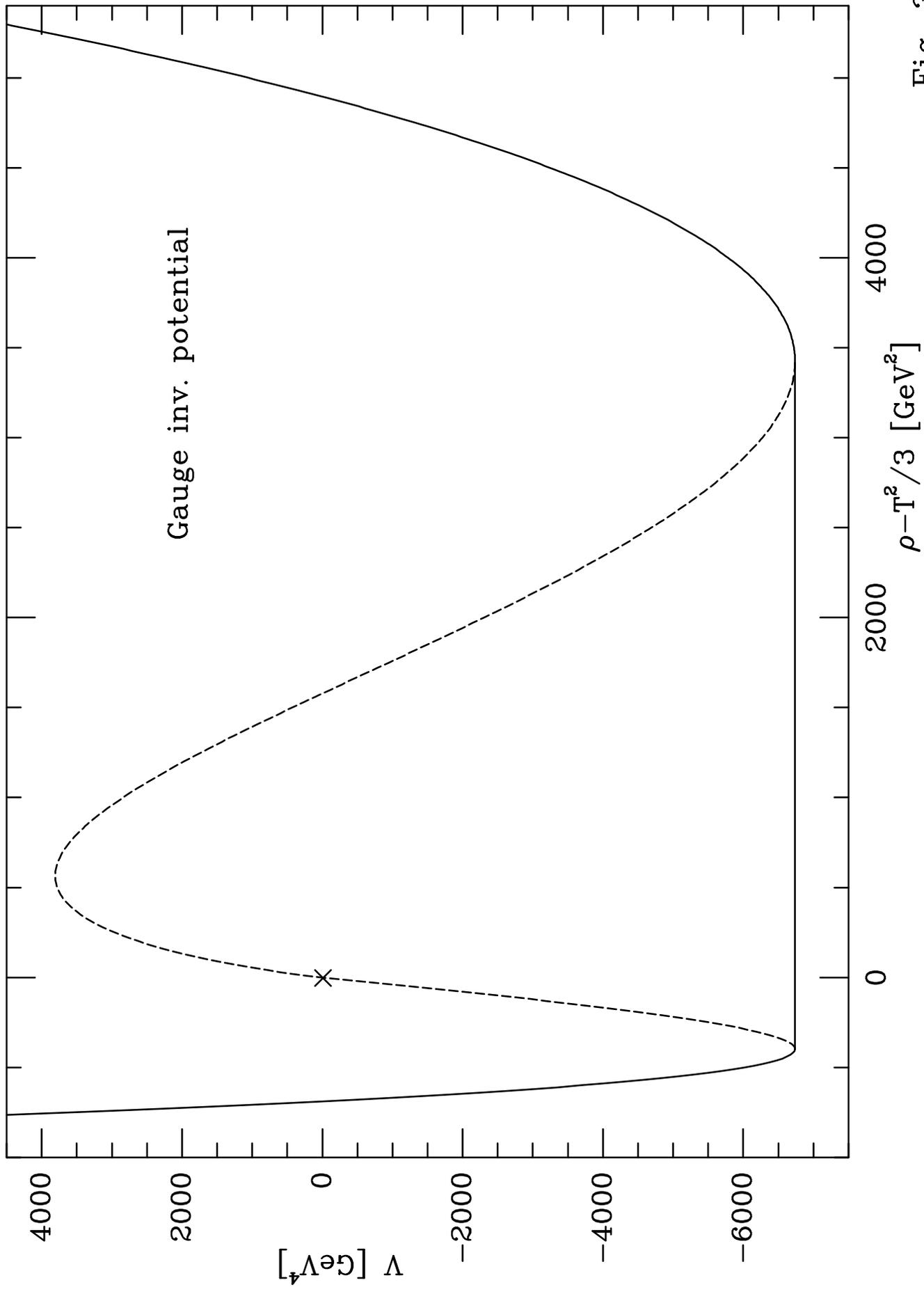
{\bf Fig.3} Gauge invariant effective potential (solid line) and its analytic
continuations from the single phase regions into the mixed phase region (dashed
line). The cross denotes the matching point.\\
\\
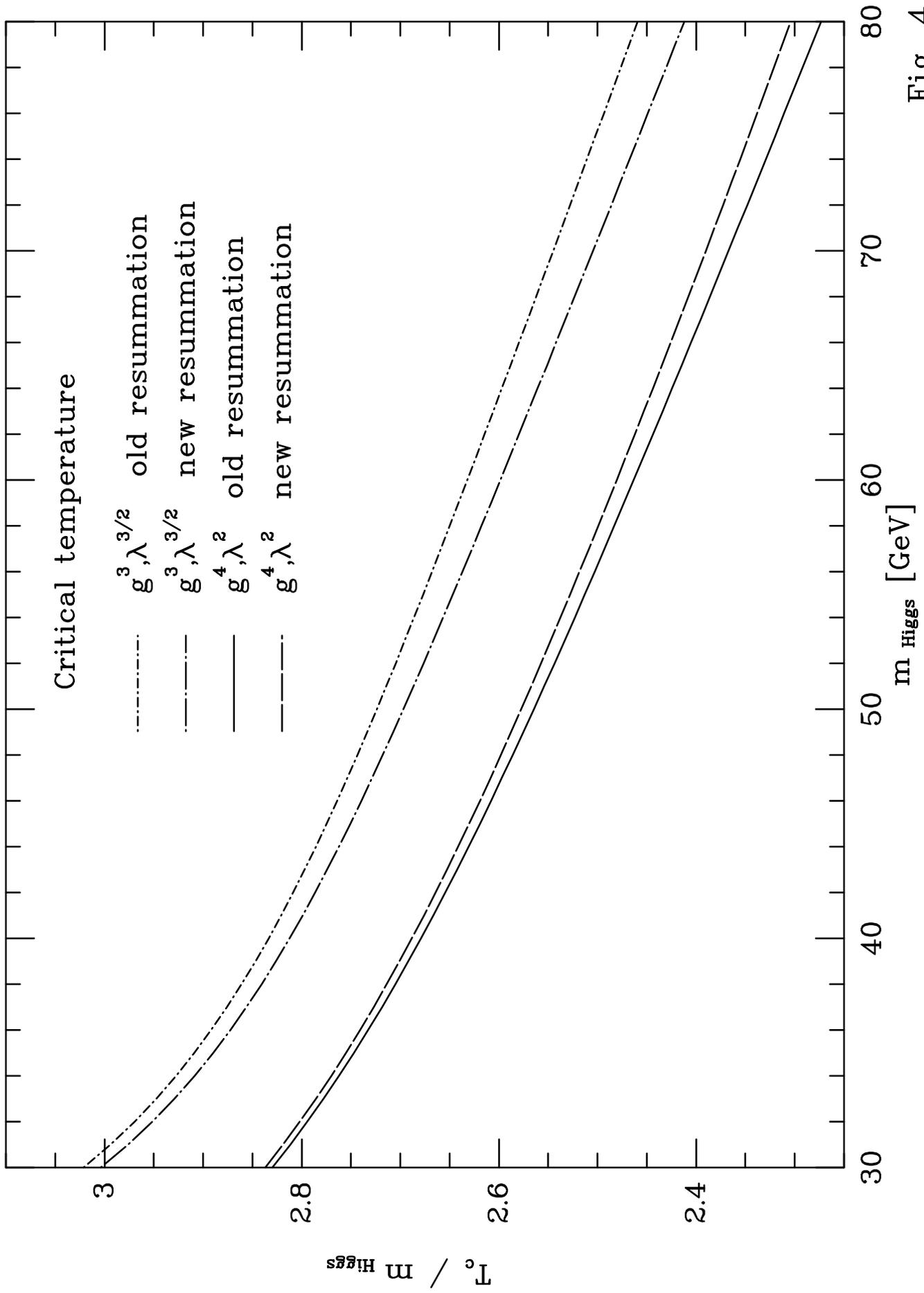
{\bf Fig.4} Ratio of critical temperature and zero-temperature Higgs mass as a
function of the Higgs mass.\\
\\
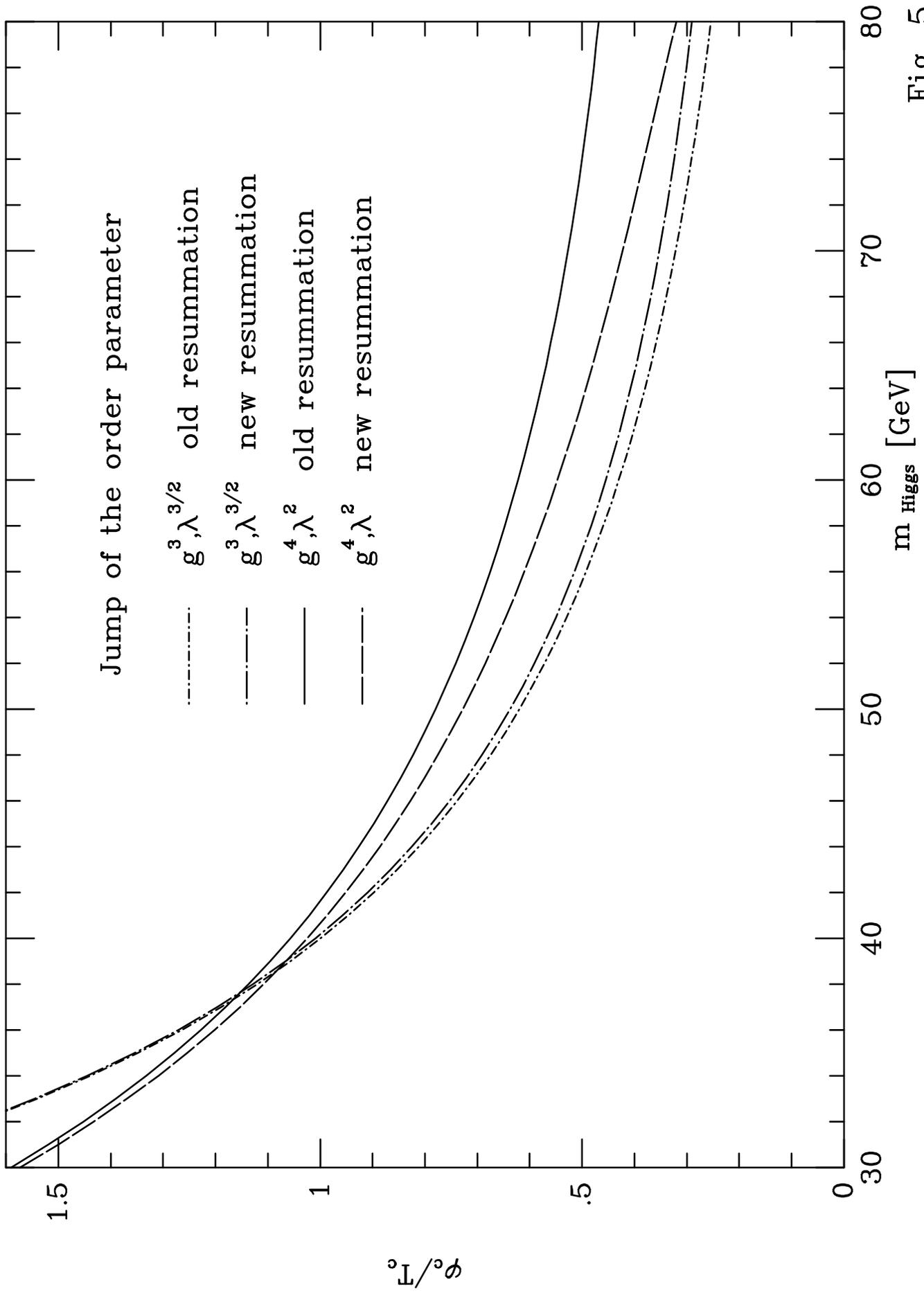
{\bf Fig.5} Square root of the jump of the order parameter $\varphi_c=
\sqrt{\Delta\rho}$ in units of the critical temperature $T_c$.\\
\\
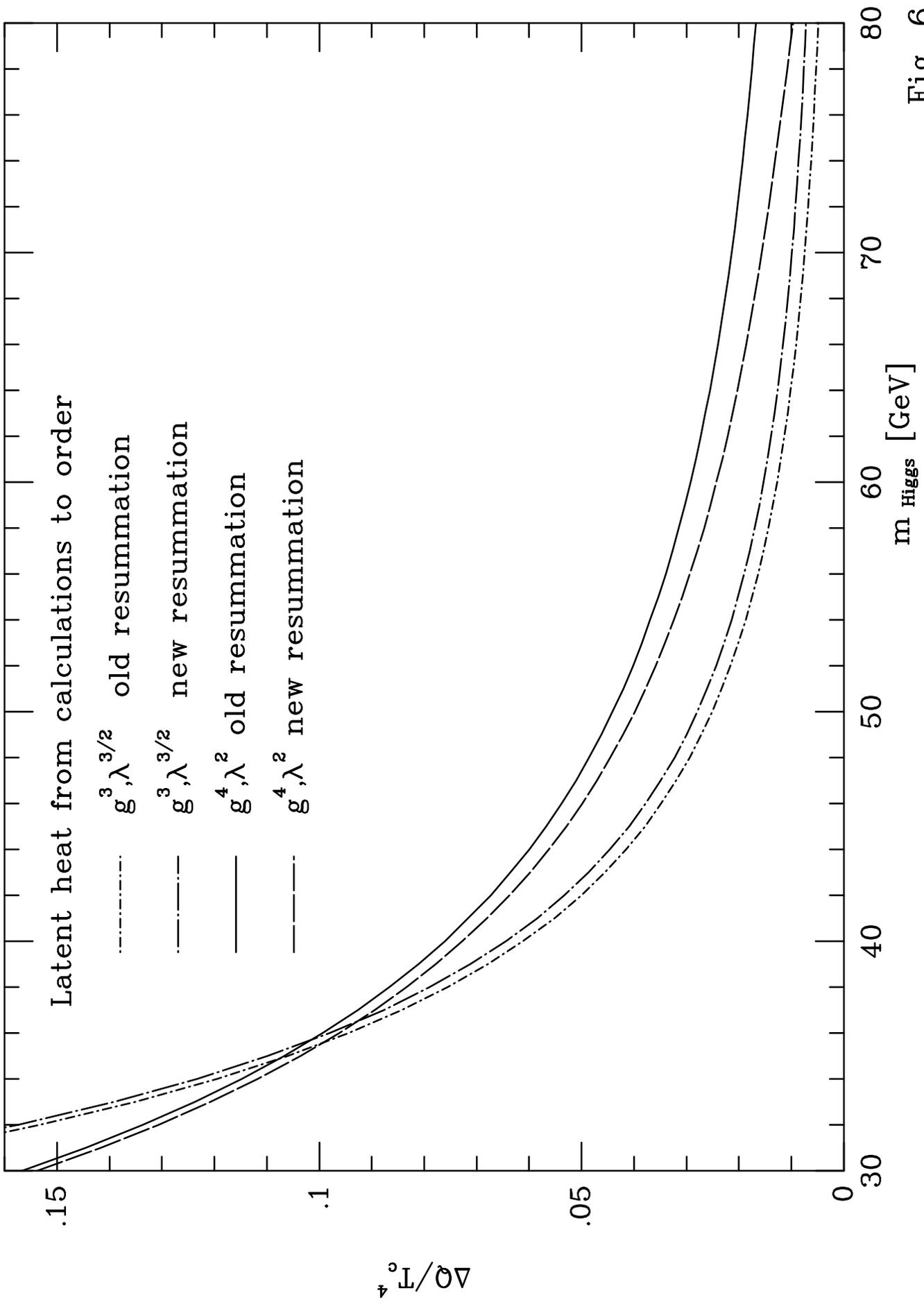
{\bf Fig.6} Higgs mass dependence of the latent heat $\Delta Q$ of the phase
transition.\\
\\
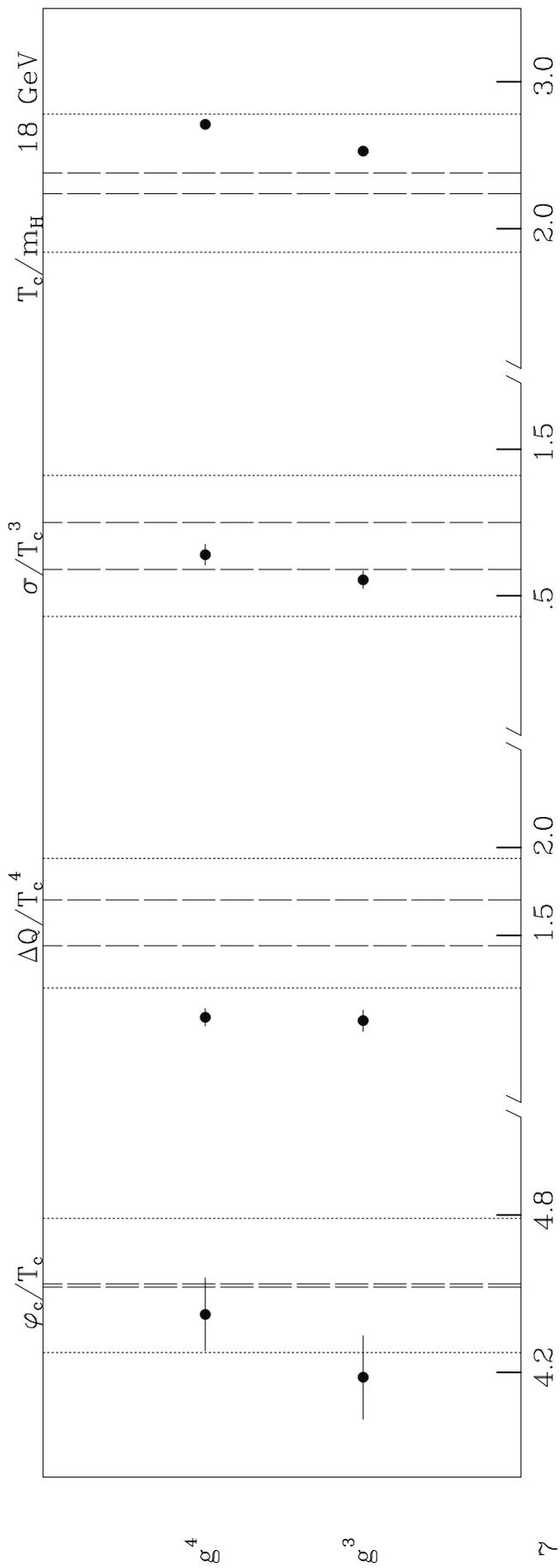
{\bf Fig.7} Comparison of the lattice data with the one-loop and two-loop
perturbative results for $m_H \approx 18$ GeV.\\
\\
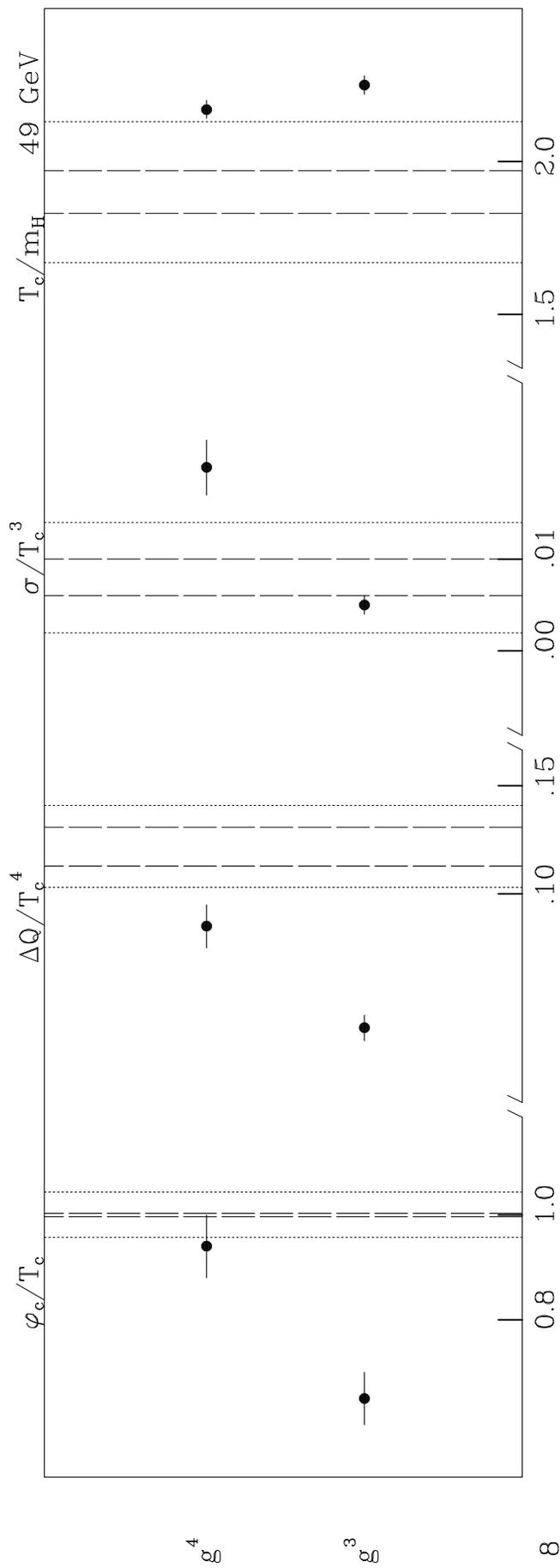
{\bf Fig.8} Comparison of the lattice data with the one-loop and two-loop
perturbative results for $m_H \approx 49$ GeV.\\
\\
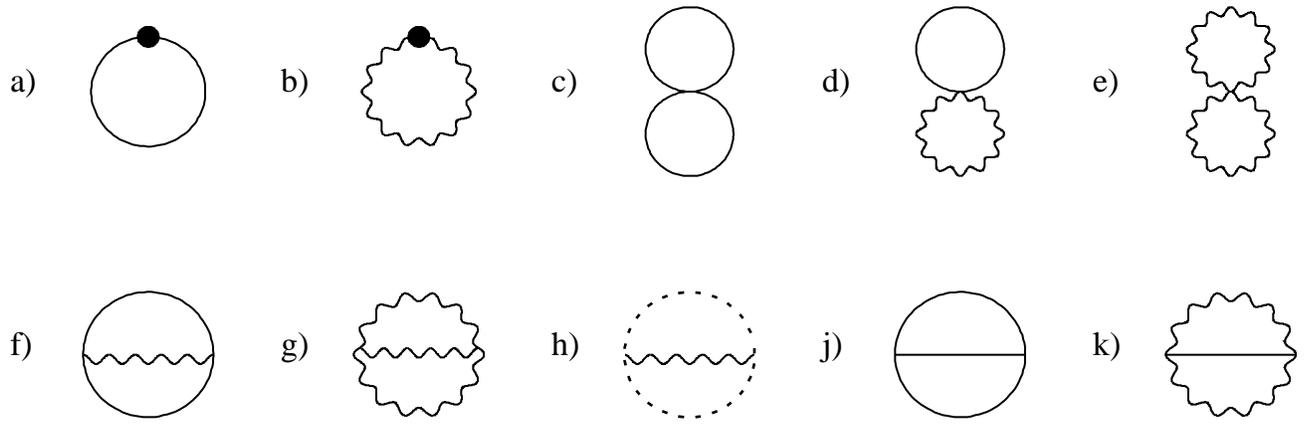
{\bf Fig.9} Two-loop diagrams contributing to the effective potential.\\
\\
\end{document}